\documentstyle[aps]{revtex}

\begin{document}

\title{Physical Reality and the Complementarity Principle}
\author{R. Srikanth\thanks{srik@iiap.ernet.in} \\
Indian Institute of Astrophysics \\
Koramangala, Bangalore- 34, India}
\maketitle
\date{}

\begin{abstract}
Consideration of the von Neumann measurement process underlying 
interference experiments shows that the uncertainty in the incoming wave,
responsible for its interference, translates during measurement into an 
uncertainty at the measuring apparatus. However, subsequent measurement on
the apparatus does not reveal any new information about the interfering wave.
This observation, in the context of recent advances in quantum 
information, suggests an argument for an
information theoretic interpretation of quantum mechanics.
\end{abstract}

~\\
In 1927 Niels Bohr propounded the complementarity principle (CP) to
rationalize the wave-particle duality of matter/energy.
According to it, the two natures pertain to the same underlying object. The
particle- and wave-manifestations are mutually 
exclusive and the particular aspect manifested depends on the experimental
set-up. This dichotomy is the price we must pay for using classical language
to describe quantum phenomena.
CP is used as an informal version of the Heisenberg Uncertainty
relation, a qualitative thumbrule to make sense of non-classical effects. 

A traditional way to illustrate wave-particle duality is via the well-known
Young's double slit experiment \cite{schiff}. Here, ``which-path" information- 
representing a particle property- and appearence of an interference
pattern on the screen- representing wave nature- are predicted by CP 
to mutually exclude each other. If we wish to avoid attributing volition to
the particle, there must exist reasons based on the phase information
accumulated by each beam as a result of interaction with the which-path
monitor that explains this duality \cite{schiff,ste90}. Zeilinger 
\cite{zeil}
has suggested that the duality might have its origin in the limitation 
imposed by the experimental set-up.
It has also been noted that the duality do not imply discrete antipodes but 
two extremes joined by
a continuum of possibilities. A further subtlety 
concerning CP is discussed in this article.

Suppose we designate by $|a\rangle$ and $|b\rangle$ the state of a particle
to be found in path $a$ and $b$ of a Young's double slit interferometer in
which, because no path detection is made, a double-slit interference is 
observed. Let $U$
be the unitary transformation that transforms the wavefunction to the
measurement space, spanned by the basis set $\{|x\rangle\}$ defined by the 
detector elements $x$, and $M$ be the measurement
whereby the particle is eventually localized at some $x$ on the screen.

Let's write $U = U_1U_2$, where $U_1$ is the unitary transformation of the
initial path states to the diffracted wavefunction basis space, and $U_2$ the 
transformation to the measurement basis space spanned by the detector elements.
Now, generally (apart from a normalization factor)
\begin{equation}
\label{u1}
|\beta\rangle \stackrel{U_1}{\longrightarrow} \sum_x 
\exp (i\theta_{\beta x})|\beta_x\rangle \hspace{1.0cm} (\beta = a, b)
\end{equation}
where $|\beta_x\rangle$ are the basis-state in the
diffracted wave moving in the direction pointing from slit $\beta$ towards
an element $x$ on the detector, and phase factors $\theta_{\beta x} = 
\theta_{\beta x}(k, d_{\beta x})$ satisfy 
$\sum_x (\theta_{ax} - \theta_{bx}) = 0$. Here
$k \equiv |{\bf k}|$ is the spatial wave
number and $d_{\beta x}$ the distance from slit
$\beta$ to element $x$. 

The mutual distinguishability of the elements of the diffracted wave basis set 
$\{|\beta_x\rangle\}$ arises from the fact that any one of them differs from 
any other in orientation and/or slit of origin.
Suppose the detector basis set is completely specified by 
$\{|0\rangle , |\phi_x\rangle\}$, where $|0\rangle$ is the initial state
of the detector and $|\phi_x\rangle$ is the detector state such that a click
occurs at element $x$. Then a mapping like $|\beta_x\rangle
\longrightarrow |\phi_x\rangle$ might be considered to specify $U_2$.
However, this is clearly not reversible-- since both $|a_x\rangle$ and 
$|b_x\rangle$ map to $|\phi_x\rangle$--, and hence cannot be unitary.
Therefore the detector basis set must be enhanced to include some internal
state of the detector element to restore unitarity. Let us say the final
internal states are $|v_a (x)\rangle$ and $|v_b(x)\rangle$. The enhanced
detector basis set is: 
 $\{|0\rangle , |\phi_x\rangle |v_a(x)\rangle , 
 |\phi_x\rangle |v_b(x)\rangle\}$. 

The transformation from the diffraction basis to detector basis for the
combined particle+detector system is written
\begin{equation}
\label{u2}
|\beta_x\rangle \otimes |0\rangle  \stackrel{U_2}{\longrightarrow} 
|x\rangle \otimes |\phi_x\rangle |v_\beta(x)\rangle \hspace{1.0cm}
(\beta = a, b)
\end{equation}
Operation $U_2$ entangles ray path and the internal mode activated.
The different transverse
momenta $p_y$ transferred by the beam on path $a$ and on path $b$ to 
the detector element at $x$ will produce different vibrations in the element.
These vibration modes are an example for valid internal states to enhance the
detector measurement space. 

The complete evolution of the particle and detector state-vector can be given 
by combining Eqs. (\ref{u1}) and (\ref{u2}):
\begin{eqnarray}
\label{uma}
\frac{1}{\sqrt{2}}\left(|a\rangle + |b\rangle \right) \otimes |0\rangle
& \stackrel{U}{\longrightarrow} &
 \sum_x |x\rangle \otimes (\exp (i\theta_{ax})|\phi_x\rangle |v_a(x)\rangle + 
\exp (i\theta_{bx})|\phi_x\rangle|v_b(x)\rangle ) \nonumber \\
& \stackrel{M}{\longrightarrow} & |x\rangle \otimes 
(\exp (i\theta_{ax})|\phi_x\rangle |v_a(x)\rangle + 
 \exp (i\theta_{bx})|\phi_x\rangle|v_b(x)\rangle ).
\end{eqnarray}
Since measurement $M$ answers the question ``where did the particle
land?", it is drawn from the set of projectors $\{|0\rangle\langle 0|, 
|\phi_x\rangle\langle \phi_x|\}$. Eq. (\ref{uma}) shows that the
amplitude contribution from both paths to the observation at $x$ results
in a superposition of vibration modes. The initial superposition leaves 
behind a remnant superposition. 
The uncertainty in path translates to uncertainty in vibrational mode because
of the entanglement of the two properties. 
Another way to understand this
result is that if the internal freedom of the detector is ignored, the
right hand side of both equations in Eq. (\ref{u2}) would have been
 identical.
This would imply that both beams $a$ and $b$ falling on $x$ produce identical
effects, in violation of the {\em classical reversibility} of Maxwell's 
equations. 

Remarkably, Eq. (\ref{uma})
 suggests that the path of the beam can be determined {\em after} the 
double-slit interference pattern has been registered, by measuring the
transverse
vibrational mode of the detector element at $x$. In the conventional
analysis, the path information is lost via omission of the detector internal
states. CP is then invoked to attribute the 
observed interference pattern to this path-indistinguishability. 

Suppose $v_a(x)$ is observed after the detector click.
This ``post-chooses" path $a$. Does this logically contradict the 
indistinguishability of paths implied by the already registered double slit 
pattern? In other words, 
does this violate the CP tenet forbidding the simultaneous
manifestation of particle and wave properties? 
Would photographs of the double-slit pattern 
dissolve if the detector is post-measured to determine the photon's path? 
The answers to these questions is ``no!", because the second measurement
(on the detector) is drawn from the set of projectors $\{|v_{\beta}(x)\rangle
\langle v_{\beta}(x)|\}$, which disentangles path from the internal mode,
as seen from Eq. (\ref{uma}).
Therefore, post-measurement of the latter cannot reveal any new information
about the incoming wave in the epoch prior to photon detection. It appears
that part of the reason we were led to the above seeming contradiction
is semantic. Current understanding,
dominated by the Copenhagen interpretation, tells us that the particle's 
position lacked reality before impinging on the screen. The internal mode
distinguishability therefore seemed to contradict the observation of
wave interference, given that the two were at first entangled.

Recent advances in quantum information, especially quantum computation,
have increasingly suggested that we can look at quantum superposition not
as compromising the reality (in the EPR \cite{epr} sense) of an
observable but rather as an important resource
for processing information. In the context of the above experiment, we reason 
from an information
theoretic standpoint that there was amplitude information from both slits prior
to and during registration on the screen. When the internal variable
is measured, amplitude information from one of the slits is (irreversibly)
lost. The possibility of post-choosing the path does not preclude interference,
since the post-choice cannot reveal any new information about the interfering
wave, in consonance with CP.
What is requisite, in the spirit of Refs. \cite{ste90}, is that the internal 
states should not introduce random phases into the wavefunction before
interference.  

~\\
I thank Dr. Unnikrishnan and Mrs. Geetha Gopakumar for discussions.

\end{document}